\documentclass[a4paper,final,12pt,onecolumn]{IEEEtran}
\usepackage{setspace}
\usepackage[cp1250]{inputenc}
\usepackage{graphicx}
\usepackage{mathrsfs}
\usepackage{url}
\usepackage{amssymb}
\usepackage{amsmath}
\usepackage{amsthm}

\newtheorem{theorem}{Theorem}

\newcommand{\matR}{{\bf R}}
\newcommand{\matV}{{\bf V}}
\newcommand{\matI}{{\bf I}}

\title{Memristors can implement fuzzy logic}

\begin{document}

\author{Martin~Klimo \IEEEmembership{Member,~IEEE}\thanks{ M. Klimo is with University of \v Zilina, \v Zilina, Slovakia}, Ondrej~\v Such\thanks{O.\v Such is with Slovak Academy of Sciences, Bansk\'a Bystrica, Slovakia}%
\thanks{Work on this paper was partially supported by the Agency of the Slovak Ministry of Education for the Structural Funds of the EU, under project ITMS:26220120007}%
}%
\maketitle

\begin{abstract}
In our work we propose implementing fuzzy logic using memristors. Min and max operations are done by antipodally configured memristor circuits that may be assembled into computational circuits.   We discuss computational power of such circuits with respect to m-efficiency and experimentally observed behavior of memristive devices. Circuits implemented with real devices are likely to  manifest learning behavior. The circuits presented in the work may be applicable for instance in fuzzy classifiers.
\end{abstract}
\begin{IEEEkeywords}
memristors, fuzzy logic, fuzzy systems, classification algorithms
\end{IEEEkeywords}

\section{Introduction}


%
{
Transistor circuits with underlying Boolean algebra operations form the basis of today's computers. Manufacturing processes for implementing them have been steadily improving in the past decades \cite{BorkharChien}. Yet the current microprocessor designs are many orders of magnitude less energy efficient than human brain \cite{fet}. The discovery of memristor \cite{TheMissingMemristorFound}, the fourth fundamental circuit element \cite{chua71}, holds a promise to narrow this gap \cite{sni11}. First of all, it can be used for high-density non-volatile memory \cite{AachenResistiveMemories09}. Secondly, boolean logic applications have been demonstrated in \cite{bor10} and \cite{AachenCRSimplication}. Specialized memristor circuits have been proposed to solve contour detection \cite{shi09} and to find the way through a maze \cite{per11}. 

Here we show (Theorem \ref{theorem:1}) that memristors quite naturally compute operations of fuzzy logic \cite{zad65}, which is an extension of classical Boolean logic. Thus memristors are capable also of processing, and memristor circuits can bridge boundaries between memory and processing units that are present in classical (von Neumann) computers. There are other approaches that promise a dramatic leap in computer performance, such as quantum computing \cite{sho97}, \cite{joh11}. However, at present memristors hold the advantage in manufacturing, being well within reach of today's nanotechnology \cite{xia09}. 
}

\section{Memristor}
\begin{figure}
\includegraphics[width=0.4\textwidth]{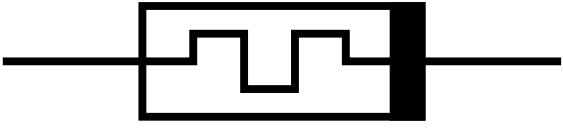}
\caption{Memristor symbol}
\label{fig:memimage}
\end{figure}

{\em Memristor} is a two-terminal circuit element similarly to resistor, capacitor and inductor.
Its existence of {\em memristor} was conjectured by L. Chua  by analogy to resistor, capacitor and inductor. It  is a two terminal circuit element, whose symbol is shown in Figure \ref{fig:memimage}. 
It obeys Ohm law relating voltage $v$ and current $i$
\begin{align*}
v = M i,
\end{align*}
where {\em memristance} $M$ is variable. The defining property of memristor is that there exists a functional relationship $g(q,\phi) = 0$ between the magnetic flux $\phi$ and charge $q$, where
\begin{align*}
q(t) = \int_{-\infty}^t i(\tau) \, d\tau,\qquad \phi(t) = \int_{-\infty}^t v(\tau) \, d \tau.
\end{align*}
In a {\em charge-controlled memristor} this relation takes the form $\phi = \phi(q)$ and by taking the time derivative one obtains
\begin{align*}
v(t) = \frac{d \phi(q)}{d t} = \frac{d \phi(q)}{d q}\frac{d q}{d t} = \frac{d \phi(q)}{d q }i(t)
\end{align*}
and thus for the charge-controlled memristor one has 
\begin{align*}
M(q) =  \frac{d \phi(q)}{d q }
\end{align*}


\section{Voltage divider circuit}

\begin{figure}[!ht]
\begin{center}
\includegraphics[width=0.6\textwidth]{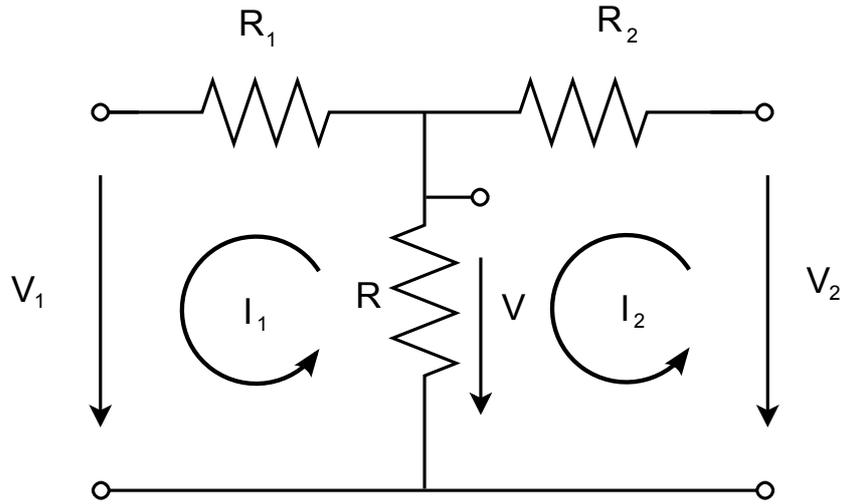}
\caption{A voltage divider}
\label{fig:diagram0}
\end{center}
\end{figure}

Memristors at infinitesimal time periods behave as resistors. In order to gain understanding of circuits in Figures \ref{fig:1}  let us first analyze the voltage divider circuit shown in Figure \ref{fig:diagram0}.
Loop current flows $I_1, I_2$ can be computed by using Kirchhoff's voltage law $\matV=\matR\, \matI$, where
\begin{align*}
\matV = \begin{pmatrix} -V_1 \\ V_2 \end{pmatrix}, \matI = \begin{pmatrix} I_1 \\ I_2 \end{pmatrix},
\matR = \begin{pmatrix} R + R_1& -R \\ -R & R+R_2\end{pmatrix},\quad 0\leq   V_1, V_2 \leq 1; R, R_1, R_2 > 0
\end{align*}
One can thus determine the currents by solving the system of linear equations, e.g. by Cramer's rule to obtain
\begin{align}
\begin{split}
I_1 &=  \frac{ -V_1(R+R_2)  + V_2R}{\Delta}, \\
I_2 &= \frac{ V_2(R+R_1)  - V_1 R}{\Delta},
\end{split}
\label{eq:i}
\end{align}
where $\Delta = (R+R_1) (R+R_2) - R^2 = R(R_1 + R_2) + R_1 R_2 > 0$.

This implies that the output voltage $V$ satisfies
\begin{align}
V = R(I_2 - I_1) = R \cdot \frac {R_1 V_2 + V_1R_2}{\Delta} = \dfrac{V_1R_2 + V_2R_1}{R_1 + R_2 + R_1R_2/R}
\label{eq:v}
\end{align}
The condition $V_1 < V$ is equivalent to 
\begin{align*}
V_1 \bigr(1 + \frac{R_2}{R}\bigl) < V_2, 
\end{align*}
thus if $R\gg \max( R_1, R_2)$ the resulting voltage $V$ is between $V_1$ and $V_2$, justifying the name ``voltage divider''.


%
\section{Ideal memristors compute min-max logic}
All memristors considered in this section are assumed to be charge driven, where memristance $M(q)$ is a {\em monotonically increasing} function satisfying
\begin{align}
\begin{split}
\lim_{q\rightarrow-\infty} M(q) &= R_{ON},\\
\lim_{q\rightarrow\infty} M(q)  &= R_{OFF},
\end{split}
\label{eq:m}
\end{align}
where $0 < R_{ON} < R_{OFF}$
We shall call such devices {\em ideal bilevel memristors}.

Fuzzy logic dates back to works of Lukasiewicz and Post. H. Weyl in 1940 proposed a fuzzy logic where propositions are assigned values in unit interval. He generalized ordinary operators as follows:
\begin{gather*}
a\text{~and~}b = \min(a,b) \\
a\text{~or~}b = \max(a,b) \\
\text{not~}a = 1 -a \\
a\text{~implies~}b = 1 -a + min(a,b) = min(1, 1-a+b),
\end{gather*}
where $a$ and $b$ take values in the interval $[0,1]$. The min-max logic has been shown to be the only choice  under natural assumptions \cite{MR0414314}. Now we will show that antipodally configured memristors   can compute the first two operations.

\begin{figure}[!ht]
\begin{center}
\includegraphics[width=0.4\textwidth]{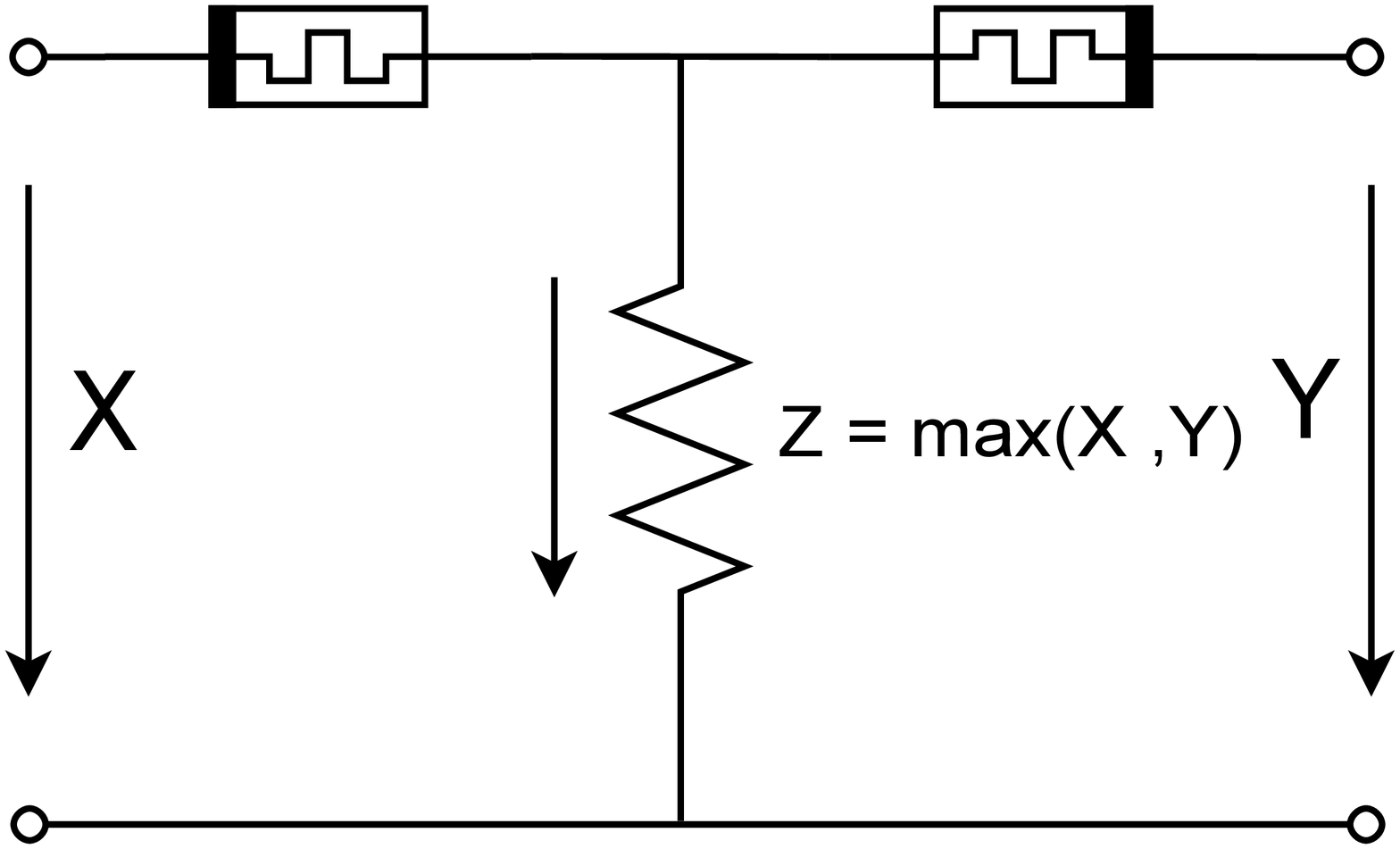}
\hfill
\includegraphics[width=0.4\textwidth]{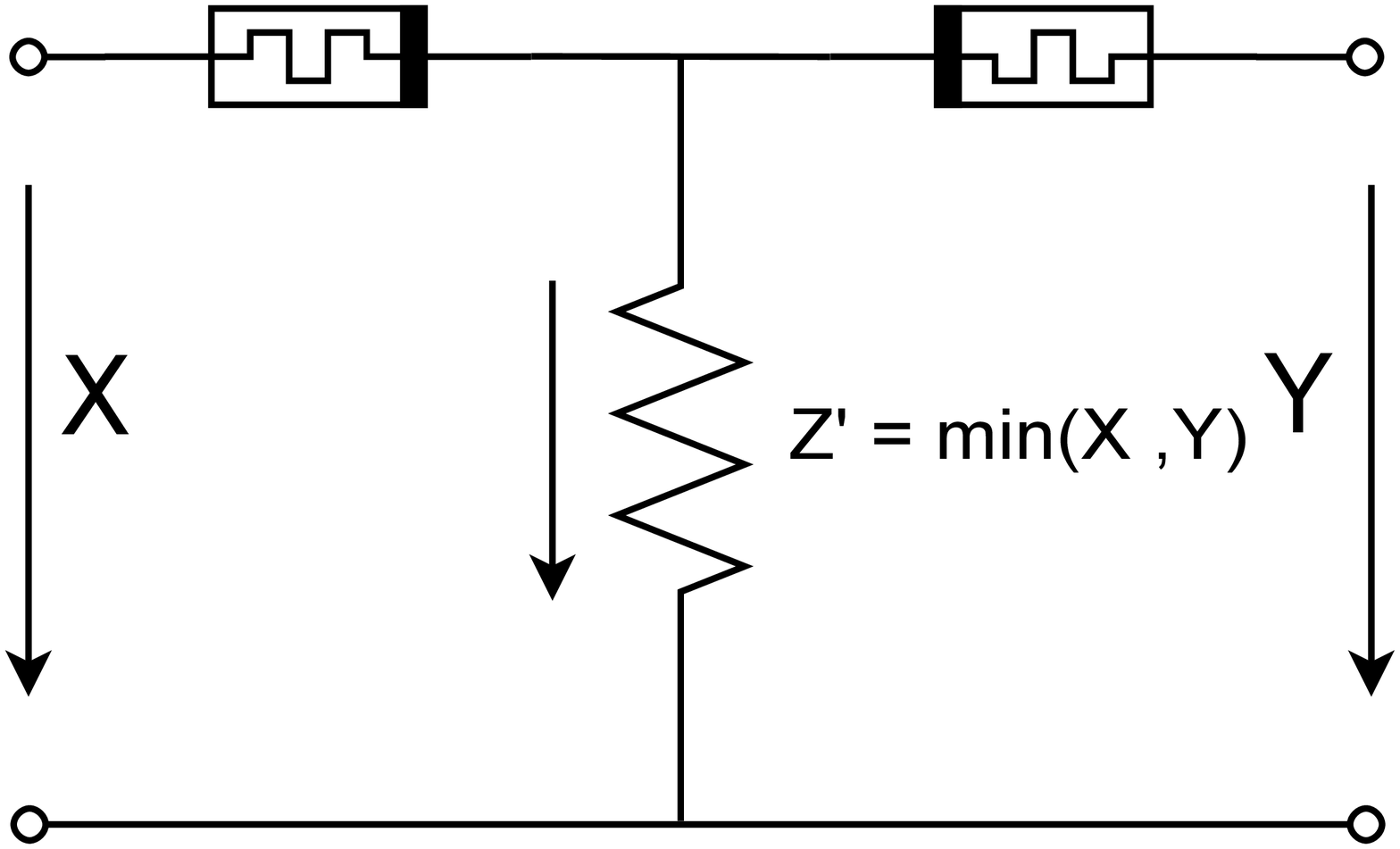}
\caption{Memristor circuits computing min and max}
\label{fig:1}
\end{center}
\end{figure}

\begin{theorem}
Suppose one sets upper pins in Figure \ref{fig:1} to constant voltages $X(t) = X$ and $Y(t) = Y$. If $R \gg R_{OFF} \gg R_{ON}$ then  
\begin{align*}
\lim_{t\rightarrow \infty} Z(t) \approx \max(X,Y)\\
\lim_{t\rightarrow \infty} Z'(t) \approx \min(X,Y)\\
\end{align*}
\label{theorem:1}
\end{theorem}
\begin{IEEEproof}
Consider the left circuit in Figure \ref{fig:1}.  Suppose $X >  (1 + R_{OFF}/R) Y$. Then from \eqref{eq:i} we see that $I_1 < 0$ and $I_2 <  0$. In fact, the currents are bounded away from $0$ by a constant $\delta > 0$ independently of time. From
the definition of ideal bilevel memristor we conclude that memristance of the left memristor approaches $R_{ON}$ and the memristance of the right memristor approaches $R_{OFF}$. Then from \eqref{eq:v} we obtain
\begin{align*}
\lim_{t\rightarrow\infty} Z(t) = \frac{X R_{ON} + Y R_{OFF}}{R_{ON} + R_{OFF} + R_{ON} R_{OFF} / R} \approx X = \max(X,Y).
\end{align*}
The symmetric case when $Y> (1 +R_{OFF}/R) X$  is handled in the same way.
In the remaining case  we have $     (1 +  R_{OFF}/R)^{-1}   \leq  X/Y \leq (1 +  R_{OFF}/R) $. Since $R_{OFF} \ll R$ we have $X \approx Y$ and thus 
\begin{align*}
\lim_{t\rightarrow\infty} Z(t) &=\lim_{t\rightarrow\infty} \frac{X M_1(t) + Y M_2(t)}{M_1(t) + M_2(t) + M_1(t) M_2(t) / R} \\
&\approx \lim_{t\rightarrow\infty} X \cdot \frac{ M_1(t) +  M_2(t)}{M_1(t) + M_2(t) + M_1(t) M_2(t) / R}  \approx X  \approx \max(X,Y),
\end{align*}
where $M_1(t), M_2(t)$ denote the memristances of the left and right memristor in time $t$. 

The proof of computation of $\min$ proceeds in a similar way.
\end{IEEEproof}

Antipodally configured memristors proved useful in other ways as well: in context of memory circuits \cite{AachenCRSmemory}, \cite{AachenCRSmemory2}, contour detection \cite{shi09}, STDP modeling \cite{synapse3} as well as logic \cite{AachenCRSimplication}. 

%
%
%


\section{Circuits using min-max operations}

\begin{figure}
\begin{center}
\includegraphics[width=0.4\textwidth]{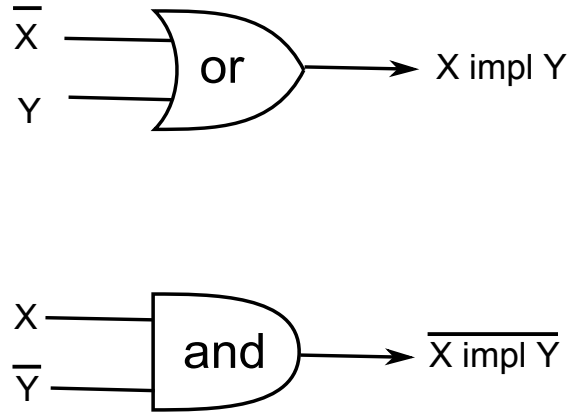}
\caption{Circuits computing implication and its negation}
\label{fig:implication}
\end{center}
\end{figure}

In order to implement a fuzzy negation or a fuzzy implication we propose to fall back to CMOS logic.  This can be done as needed, or by precomputing negations of variables at the beginning of a computational circuit \cite{blum}. Figure \ref{fig:implication} shows how for a given set of inputs and their negations one can compute implication function and its negation. 
A CMOL framework for interface
between nanoscale memristor crossbars and traditional CMOS components have been suggested in  \cite{likharevCMOL}. 

However, even without these operations, min-max circuits can carry out interesting computations. For instance the bitonic sorting algorithm\cite{bat68} uses a feed-forward network consisting of circuits computing min and max values. It is able to sort $N$ elements in $O(\log^2(N))$  parallel steps. Thus min-max circuits can compute robust statistics of their inputs, such as the median. In particular, median computation can be used for  majority vote of a set of binary fuzzy classifiers in the manner of random forests \cite{FuzzyDecisionTrees1}, \cite[Chapter 15.3]{esl}.

\newcommand{\meff}{\mu_{\text{eff}}}

One caveat is that the circuits listed in Figure \ref{fig:1} compute min and max only approximately. The approximation errors are controlled by the ratio $\meff = R_{OFF} /R_{ON}$, that we shall call {\em m-efficiency}. Experimental devices have been constructed with m-efficiency between 10 and $10^{6}$ \cite{likharevCMOL}.   If the  m-efficiency is not large enough, the output of circuits in Figure \ref{fig:1} is
\begin{align*}
\lim_{t\rightarrow \infty} Z(t) &= \frac{\max(X,Y)  + \meff^{-1} \min(X,Y)}{1 + \meff^{-1}} \\
\lim_{t\rightarrow \infty} Z'(t) &= \frac{\min(X,Y)  + \meff^{-1} \max(X,Y)}{1 + \meff^{-1}} 
\end{align*}
Figure 3 shows the effect of varying m-efficiency on the output from a bitonic sorting network. The picture demonstrates a strong linearization effect of a bitonic network for low values of m-efficiency whose impact fades with increasing m-efficiency.

\begin{figure}
\begin{center}
\includegraphics[width=0.6\textwidth,angle=270]{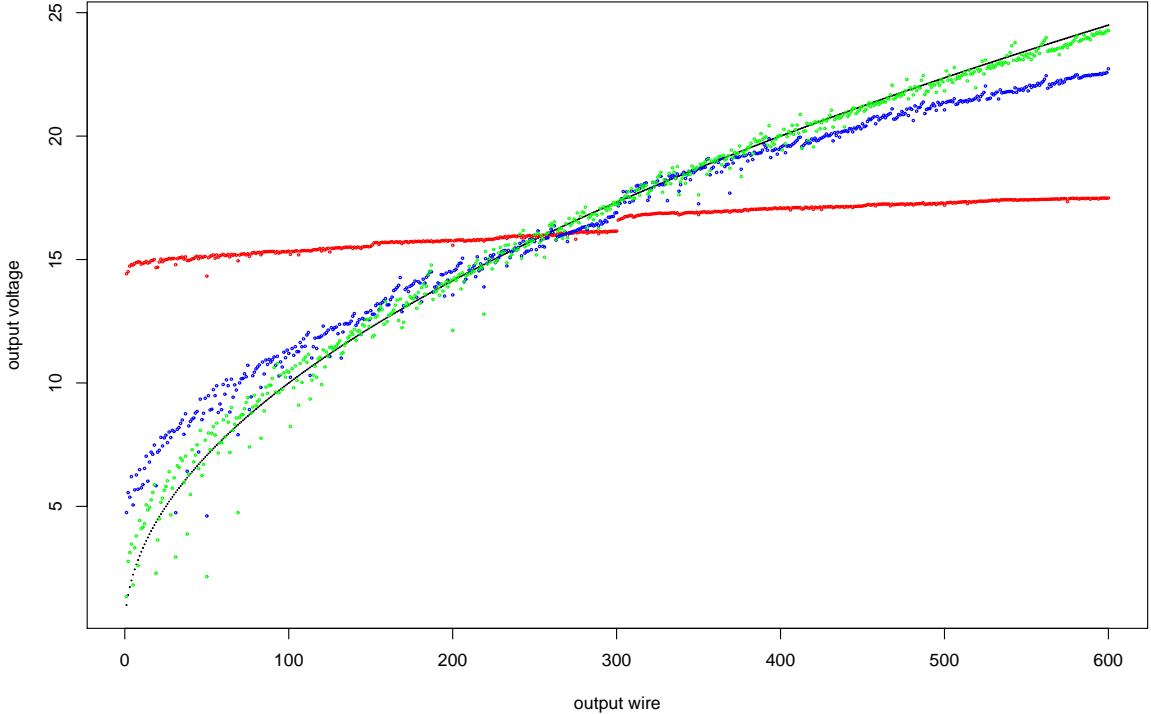}
\caption{Simulation of output of bitonic sorting network implemented using memristors. The black curve indicates the output of an ideal bitonic network, red, blue, green points correspond to m-efficiency values 10, 100, 1000. The original data consisted of randomly shuffled square roots from 1 to $\sqrt{600}$.}
\label{fig:3}
\end{center}
\end{figure}

Only partial results are known for the behavior of recursive fuzzy logic circuits. Chaotic behavior of self-referential fuzzy systems is shown in \cite{grim93}. A well known liar's paradox is proved to have a solution under quite general conditions in \cite{keh}. 

\section{Computation with non-ideal devices}
Let us compare the behavior of an ideal bilevel memristor with the behavior of experimentally studied devices, especially $TiO_2$ memristors. Bilevel resistance is indeed typical for metal oxide devices. (Notable is work of W. Lu \cite{LuSynapse}, who showed that  amorphous silicon memristor is capable of maintaining many states between $R_{ON}$ and $R_{OFF}$).
Monoticity assumptions are mildly violated by $TiO_2$ memristors, perhaps due to multiple mechanisms that play role in their memristive behavior. Better monoticity can be seen for instance in $TaOx$ memristors \cite[Fig.3]{HighEndurance} in which simpler mechanisms are expected.

Perhaps the biggest concern raised by experimental data is the exponential dependence of switching time on the applied voltage as seen for instance in \cite[Figure 1b), d)]{HPlognormalSwitching}. The figure for instance indicates that it takes $\approx 10^5$ longer to switch $TiO_2$ device with half voltage. If the memristance of the device during that time stayed in the interval $[R_{ON}, R_{OFF}]$ then charge that flows under voltage $V$ in time interval $\Delta t$ is between $V \Delta t/ R_{OFF}$ and $V\Delta t / R_{ON}$.  Thus these data would imply that m-efficiency of the charge-driven memristor would have to be $ \geq \frac12 10^5$. The paper does not indicate m-efficiency of the device, but this would be outside of the typical range for a $TiO_2$ device\cite{HighEndurance}. A violation of memristor behavior does not necessarily preclude applicability of Theorem 1. For its conclusion to hold, one only needs a weaker condition, namely
\begin{itemize}
\item[(*)]  current/voltage bounded away from zero eventually switches memristance of the device to $R_{OFF}$ or $R_{ON}$  depending on its sign
\end{itemize}
This condition is explicitly stated to hold for $TiO_2$ memristors in the seminal paper \cite[pg. 82]{TheMissingMemristorFound}. 

For practical purposes the exponential switching time means that short application of voltages to circuits in Figure \ref{fig:1} do not necessarily compute $\min$ and $\max$ in real time. They compute them, if the previous computation yielded min and max, and the order of operands is again the same. If however the order of operands changes and their difference is small, the memristances may change only partially. Precise dynamics are highly nonlinear \cite{wongCRS}. Only repeated applications of  the same inputs will eventually cause the circuit to compute the correct result. This can be viewed as a learning process of the fuzzy circuit.

Memristive devices are often  driven by voltage pulses above ``threshold voltage''. This is an empirically determined amplitude that is known to switch the state of a memristive device (or a complementary resistive switch \cite{AachenCRSmemory2}). This approach is however more suitable for discrete levels of circuit inputs, such as those that appear in classical (non-fuzzy) random forest classifiers.

\section{Conclusion}
We have shown that properties of memristors can be applied to voltage based fuzzy logic computations. In the near future 
industry is expected to deliver first commercial circuits using memristors.  Memristor circuits may prove
to be ideally suited for low power memory storage, signal processing, pattern recognition and other applications. Memristor fuzzy logic circuits complement other suggested uses of memristors, such as CMOL, reconfigurable CMOS circuits, CrossNets   and other neural networks \cite{likharevCMOL}, \cite{likharevCrossNets}.

There is another way to look at the results presented in  our work. It has been observed that memristive phenomena closely mimic behavior of synapses \cite{LuSynapse},\cite{V1memristor}, \cite{synapse3}. Since fuzzy logic arises in quite simple context among memristor circuits, it suggests itself for simulations of behavior of complex biological neural systems.

\bibliographystyle{elsarticle-num}
\bibliography{my_bib}
\end{document}